\documentclass[aps,twocolumn,amsmath,amssymb]{revtex4}
\usepackage{graphicx}
\usepackage{dcolumn}
\usepackage{bm}

\begin{document}

\title{Kondo Effect in a Many-Electron Quantum Ring}

\author{A. Fuhrer,$^1$ T. Ihn,$^1$ K. Ensslin,$^1$ W. Wegscheider,$^2$ M. Bichler$^3$}

\affiliation{
$^1$Solid State Physics Laboratory, ETH Z\"urich, 8093 Z\"urich, Switzerland\\
$^2$Institut f\"ur experimentelle und angewandte Physik, Universit\"at Regensburg, Germany\\
$^3$Walter Schottky Institut, Technische Universit\"at M\"unchen, Germany}

\date{\today}

\begin{abstract}
The Kondo effect is investigated in a many--electron quantum ring as a function of magnetic field. For fields applied perpendicular to the plane of the ring a modulation of the Kondo effect with the Aharonov--Bohm period is observed. This effect is discussed in terms of the energy spectrum of the ring and the parametrically changing tunnel coupling. In addition, we use gate voltages to modify the ground--state spin of the ring. The observed splitting of the Kondo--related zero--bias anomaly in this configuration is tuned with an in--plane magnetic field.  
\end{abstract}

%\pacs{Valid PACS appear here}

%\keywords{Suggested keywords}%Use showkeys class option if keyword display desired
\maketitle

%\section{\label{sec:level1}Introduction}
In quantum dots strongly coupled to leads the Kondo--effect~\cite{98Agoldhaber,98cronenwett,98schmid} emerges as an enhanced zero--bias conductance in the Coulomb--blockade if the dot has non--zero total spin. The effect can be attributed to the formation of a coherent spin state between the electrons in the leads and on the quantum dot.
The relevant energy scale is given by the Kondo temperature $T_\mathrm{K}$~\cite{78haldane}. For temperatures much larger than $T_\mathrm{K}$ the enhanced conductance disappears. The same is true if the source--drain bias is increased above $k_BT_\mathrm{K}$ leading to the characteristic zero--bias anomaly (ZBA) in Coulomb--blockade diamonds, i.e., a line of enhanced conductance in the Coulomb--blockade valley around zero bias.  A magnetic field can be employed to lift the spin degeneracy of the unpaired spin via the Zeeman splitting~\cite{98Agoldhaber,98cronenwett,98schmid}. Gate voltages allow to change between the empty orbital, Kondo, and mixed valence regimes~\cite{98Bgoldhaber} and to tune the Kondo temperature $T_\mathrm{K}$~\cite{98cronenwett,00wiel}. In the case of consecutive filling of each orbital with spin--up and --down, it is expected that the Kondo effect occurs for an odd number of electrons where one spin is unpaired. In the
absence of this odd--even behavior~\cite{00schmid}, it becomes necessary to consider dot states with spin $s_N>1/2$. Close to an orbital crossing the occurrence of singlet--triplet transitions leads to new variants of the Kondo effect~\cite{00nygard,02wiel,02kyriakidis}. Similarly, double quantum dots have been employed to investigate the coupling of two dot spins~\cite{04craig,04chen,01jeong} where the transition from Kondo--like to antiferromagnetic coupling is of interest. These experiments have also spurred a number of recent theoretical studies on novel features of the Kondo effect in quantum dots~\cite{02hofstetter,02boese,01pustlinik,01pohjola,03pustilnik}. The Kondo effect was also observed in electronic transport through quantum dots defined by carbon nanotubes~\cite{00nygard}, single molecules~\cite{02liang} and single atoms~\cite{02park}.

In the present paper we investigate the Kondo effect in a ring--shaped quantum dot. Previous experiments on mesoscopic rings include the measurement of the transmission phase of a Kondo correlated quantum dot embedded in a ring~\cite{02ji}.  Keyser et al.~\cite{03Akeyser} investigated the Kondo effect in a few--electron quantum ring as a function of magnetic field applied perpendicular to the sample. They found a periodic modulation of the Kondo effect with a fraction of the Aharonov--Bohm (AB) period and linked this behavior to interaction driven localization of the electrons along the ring for small electron numbers. 
Here we investigate a many--electron ring and find a modulation of the Kondo effect with a period determined by the addition of single flux quanta through the ring. In addition, we show that the occurrence of a ZBA in successive Coulomb diamonds can be linked to a triplet state which we can tune through a triplet--singlet transition using asymmetric gate voltage~\cite{03fuhrer} and an in--plane magnetic field. 

%\section{\label{sec:level2}Experimental results}
\begin{figure}[htb]
\centering
\parbox[b]{8.5cm}{
\includegraphics[width=8.5cm]{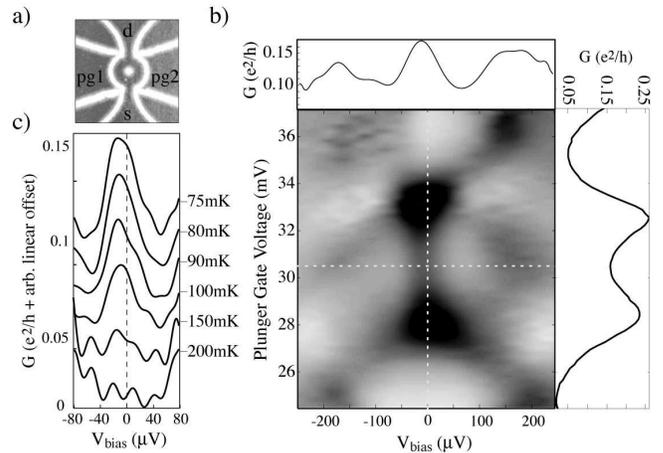}
}
\parbox[b]{8.5cm}{
  \caption{\small (a) SFM micrograph of the quantum ring structure. (b) Grayscale image of the differential conductance through the ring from source 's' to drain 'd' as a function of DC--bias and gate voltage applied to both plunger gates 'pg1' and 'pg2'. Cross--sections along the white dashed lines show the zero bias anomaly (ZBA) as a function of bias (top) and the enhanced valley conductance (right) indicative of the Kondo effect. (c) ZBA for various temperatures. The Kondo effect vanishes for temperatures above 0.2K. }\label{fig1}}
\end{figure}
The quantum ring sample was fabricated using the tip of a scanning force microscope (SFM) to oxidize locally the surface of an [Al]GaAs heterostructure containing a two--dimensional electron gas (2DEG) 34 nm below the surface~\cite{03fuhrer,02Afuhrer}. An SFM--micrograph of the ring structure is shown in Fig.~\ref{fig1}(a). The bright oxide lines lead to insulating barriers in the 2DEG. The ring has an average radius $r_0=132~\mathrm{nm}$ and is connected to source (s) and drain (d) through two tunnel barriers. The number of electrons in the ring is tuned by applying gate voltages $V_\mathrm{pg1}=V_\mathrm{pg2}$ to the two in--plane plunger gates pg1 and pg2.  A homogeneous metallic top gate was used to tune the ring into the strong coupling regime, which is in contrast to previous experiments~\cite{03fuhrer,01Bfuhrer} in which the same ring structure was measured with weak coupling. The top gate voltage is kept  fixed at $V_\mathrm{tg}=222.7$~mV for all measurements shown in this paper. For this configuration we estimate an average charging energy $E_\mathrm{C}\approx200~\mu$eV from Coulomb diamond measurements. The amplitude of the largest conductance resonances was of the order of $0.3e^2/h$ and we estimate the coupling $h\Gamma = 100\pm50~\mu$eV from the width of the conductance peaks. The measurements were performed with DC techniques in a dilution refrigerator with a base temperature of 50\,mK. Values for the differential conductance through the ring were obtained by taking numerical derivatives of DC measurements as a function of source--drain bias.

\begin{figure}[htb]
\centering
\parbox[b]{8.5cm}{
\includegraphics[width=8.5cm]{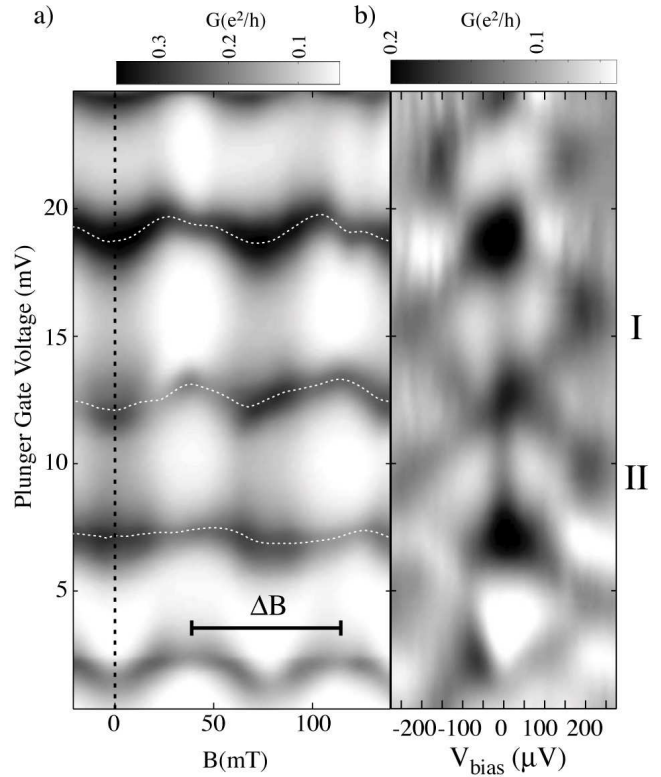}
}
\parbox[b]{8.5cm}{
  \caption{\small (a) Conductance oscillations as a function of gate voltage and magnetic field applied perpendicular to the sample. (b) Coulomb diamonds measured along the dashed black line in (a). The two diamonds (marked I and II) show a pronounced zero bias anomaly.}\label{fig2}}
\end{figure}

Figure~\ref{fig1}(b) shows parts of three Coulomb diamonds in a grayscale plot of the differential conductance as a function of plunger gate voltage and DC--bias voltage. In the Coulomb--blockaded region of the central diamond ($V_\mathrm{pg}\approx30~\mathrm{mV}$) we observe the ZBA which is characteristic for the Kondo effect. This is more clearly visible in the cross--sections along the white dashed lines: the trace to the right shows the enhanced conductance in the central Coulomb valley between the Coulomb resonances. The top trace shows a cut through the ZBA in the middle of the diamond.  From the decay of the ZBA with increasing temperature [Fig.\ref{fig1}(c)] we estimate the Kondo temperature $T_\mathrm{K}$ to be about 200~mK~\cite{00wiel}. In agreement with the expected odd--even behavior for successive filling of orbitals~\cite{01Bfuhrer}, the top and bottom diamonds in Fig.~\ref{fig1}(b) do not show a zero bias anomaly. However, we also observe cases where successive diamonds show a ZBA [see Fig.~\ref{fig2}(b) for an example]. This is in agreement with previous experiments on this structure where we show on the one hand that due to the screening of the metallic top gate interactions are relatively small and many spin pairs are observed~\cite{01Bfuhrer,03fuhrer}. On the other hand, we have also shown that, while ground state spins $s_N\geq3/2$ are rare, triplet states ($s_N=1$) occur rather frequently and can be tuned with gate voltages~\cite{03fuhrer,04ihn}. Therefore, the occurrence of  ZBA's in successive diamonds in the Kondo regime are plausible.

The energy spectrum of a quantum ring as a function of magnetic field applied perpendicular to the plane of the ring was measured in this ring structure for the weak coupling regime~\cite{01Bfuhrer}. It was found that both the broadening and the energy of the resonant level are modulated with the AB period. Figure~\ref{fig2}(a) shows that the periodic modulation of Coulomb peak position and amplitude persists in the strong coupling regime. Conductance resonances show up in dark gray and positions of  maxima are indicated by white dashed lines. The period $\Delta B=75~\mathrm{mT}$ corresponds to adding a single flux quantum to the average ring area. 

We find that the conductance in the Coulomb valley between the peaks also oscillates with the AB period. The grayscale plot of the differential conductance as a function of bias and gate voltage at B=0\,T in Fig.~\ref{fig2}(b) indicates that the enhanced conductance between the peaks is due to the Kondo effect. The corresponding ZBA is strongest for the second Coulomb diamond from the bottom (marked II) in agreement with Fig.~\ref{fig2}(a) where this valley shows the largest conductance at zero field.  The modulation of the enhanced conductance suggests that the strength of the Kondo effect is modulated by the flux penetrating the ring.

\begin{figure}[htb]
\centering
\includegraphics[width=3.4in]{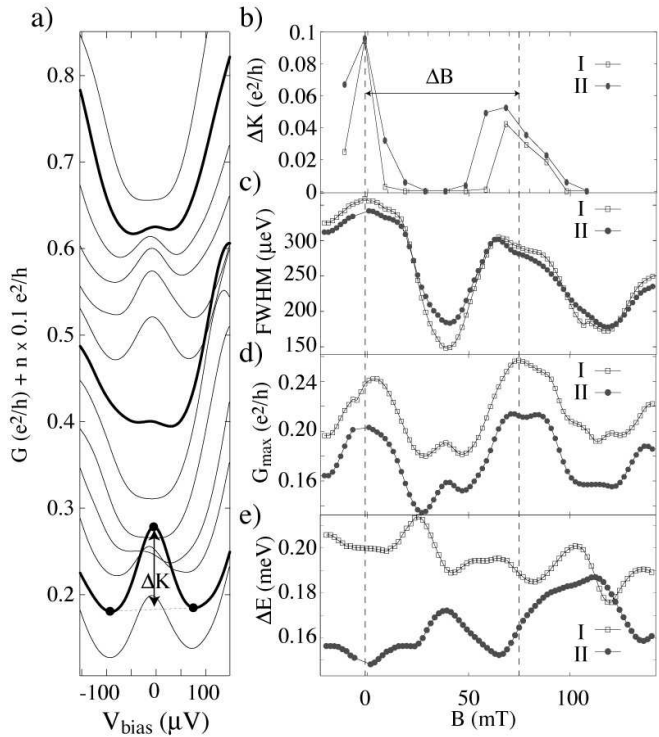}
  \caption{\small (a) Zero bias anomaly for various perpendicular magnetic fields B. The bold black lines are for field values $B = 0~mT, 50~mT, 100~mT$ from bottom to top. Curves are offset for clarity. The strength of the Kondo effect is estimated from the peak to valley conductance $\Delta K$ inidcated by the arrow. (b) $\Delta K$ for the two Coulomb diamonds I and II as a function of B. Average width FWHM (c), amplitude $G_\mathrm{max}$ (d) and level separation $\Delta E$ (e), of the conductance peaks bordering Coulomb diamond I and II respectively.}\label{fig3}
\end{figure}

We verified this statement by measuring the two Coulomb diamonds (marked I and II) as a function of perpendicular magnetic field in increments of 10~mT. From these measurements we extract the differential conductance traces in the center of diamond II as depicted in Fig.~\ref{fig3}(a).  Each trace has been vertically offset by $0.1e^2/h$ for clarity and the bold lines mark the traces at $B = 0~\mathrm{mT}, 50~\mathrm{mT}, 100~\mathrm{mT}$ (from bottom to top). The amplitude of the ZBA is strongest at zero magnetic field, decreases until it vanishes completely then increases again to reach a second maximum after one AB--period.

For a more quantitative analysis we determine the value $\Delta K$ which is the magnitude of the ZBA [see arrow and black dots in Fig.~\ref{fig3}(a) at B=0~mT]. Figure~\ref{fig3}(b) shows the magnetic field dependence of this value for the two diamonds I and II. For both traces we observe the suppression and recurrence of the ZBA after a flux quantum is added to the ring area. We did not measure the full diamonds for fields larger than $B$=110~mT but found the background conductance between the peaks to oscillate with the AB--period up to at least $B$=400~mT. 

%\section{\label{sec:level3}Discussion}
In quantum dots the Kondo state is well developed for temperatures below~\cite{78haldane}:
\begin{equation}
T_\mathrm{K}\sim\frac{\sqrt{h\Gamma U}}{2 k_B}\cdot e^{-\frac{\pi\left(\epsilon_F-\epsilon_0\right)\left(\epsilon_F-(\epsilon_0+U)\right)}{h\Gamma U}}\label{kondot}\nonumber
\end{equation}
where $U$ is the on site Coulomb repulsion, $\Gamma$ denotes the tunnel coupling to the leads and $\epsilon_0$ is the energy of the singly occupied orbital below the Fermi energy $\epsilon_F$ in the leads. The model assumes a single orbital that can be occupied with 0,1 or 2 electrons. In the center between two Coulomb peaks $T_\mathrm{K}\sim e^{-\pi U/4 h \Gamma}$ depends exponentially on the coupling $\Gamma$ and on the addition energy U. In order to estimate the coupling in our experimental data we fit all the Coulomb peaks in Fig.~\ref{fig2}(a) at all magnetic field values. This procedure gives both the FWHM and the amplitude $G_\mathrm{max}$. For a particular Kondo valley the parameters of the neighboring conductance peaks are averaged. The dependence of the resulting parameters on magnetic field is shown in Figs.~\ref{fig3}(c) and (d). 
Both values the FWHM and  $G_\mathrm{max}$ show a clear AB--periodicity and are large at $B=0\mathrm{T}$. This allows the interpretation that $T_\mathrm{K}$ is tuned through the periodic modulation of the coupling, which leads to the periodic occurrence of the Kondo effect in a magnetic field. A similar AB periodic modulation of the Kondo effect was observed in quantum dots in the quantum Hall effect regime, where the different coupling of Landau levels to the leads and their flux dependent degeneracy is given as an explanation of the effect~\cite{02fuhner,03stopa,01keller}.

A more refined model of the magnetic field dependence would need to take into account that the addition energy $U$ changes as the magnetic field is tuned. We expect that the Kondo effect would be suppressed with increasing $U$. In Fig.~\ref{fig3}(e) we plot the peak separation $\Delta E$, which we take as an estimate of $U$, for each Kondo valley as a function of magnetic field. The corresponding units are determined from measurements of the gate lever arm~\cite{03fuhrer}. Comparing with the behavior of $\Delta K$ [Fig.~\ref{fig3}(b)] we find no clear correlation to the level separation. Together with the fact that the modulation of $\Delta E$ is less than 10\% of the total level separation we conclude that its influence on the modulation strength of the Kondo effect is small.

\begin{figure}[htb]
\centering
\includegraphics[width=3.4in]{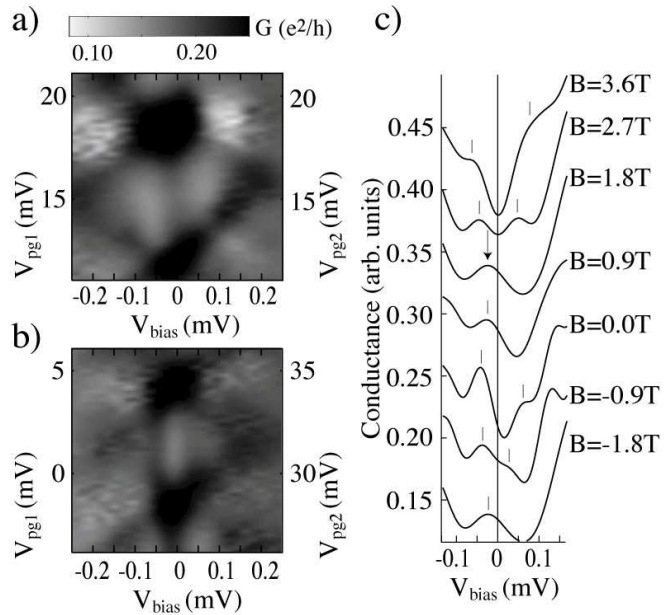}
  \caption{\small (a) Coulomb diamond at $B=0$~mT with ZBA for symmetric gate voltages $V_\mathrm{pg1}$ (left figure axis) and $V_\mathrm{pg2}$ (right figure axis)  (b) The same diamond when $V_\mathrm{pg1(pg2)}$ is raised(lowered) by 15~mV. The conductance peak at zero bias is split with maxima at $V_\mathrm{bias}\approx\pm50~\mu eV$. (c) Splitt conductance peak as a function of in--plane magnetic field. The single peak is recovered at a magnetic field value of $B=1.8~\mathrm{T}$}\label{fig4}
\end{figure}
Having established the AB--periodicity of the ZBAs we now discuss that the two ZBA's under consideration occur in successive diamonds. Such observations have been attributed to spin states of quantum dots with $s_N>1/2$~\cite{00schmid}. In our case we know from measurements in the weak coupling regime~\cite{03fuhrer,04ihn} that orbital levels are not always filled successively. In particular, we can induce orbital crossings between states that are extended around the ring and states localized in one of the rings arms by changing the two gate voltages $V_\mathrm{pg1}$ and $V_\mathrm{pg2}$ asymmetrically. Due to the difference in the Hartree energies of these orbitals they are no longer successively filled with spin--up and --down. This leads to the observation of a singlet--triplet transition even though the exchange energy $\approx 25~\mu\mathrm{eV}$ was found to be small. 
In the following we use the same procedure for tuning a singlet--triplet transition in the strong coupling regime. Figure~\ref{fig4}(a) shows diamond I for symmetrically applied gate voltages (see left and right axis of the figure). When we decrease (increase) the voltage $V_\mathrm{pg1(pg2)}$ applied to the left (right) plunger gate respectively [see Fig.~\ref{fig4}(b)] the ZBA splits into two conductance peaks at finite bias. We can tune the splitting by applying an in--plane magnetic field as indicated in Fig.~\ref{fig4}(c). The lowest trace shows a small ZBA at $B_\mathrm{||}=-1.8~\mathrm{T}$ which splits with increasing $B_\mathrm{||}$ by about $100\mu \mathrm{eV}$ at zero field. When $B_\mathrm{||}$ is further increased the splitting disappears again at  $B_\mathrm{||}=+1.8~\mathrm{T}$ and reappears for even larger fields. 
Such splittings have been attributed to a singlet ground--state with a triplet excited state in previous experiments~\cite{00nygard,03kogan,02wiel}. Under this assumption we can extract a $g$--factor $g^*=0.49\pm0.08$ which is consistent with the bulk $g$--factor of GaAs. It remains to be investigated what differences arise in a situation where the transition from a doubly occupied orbital to two singly occupied orbitals is driven by fluctuations in the Hartree interaction rather than by the exchange interaction~\cite{03fuhrer}. 

In conclusion, we have shown that the Kondo effect in a quantum ring is periodically modulated with the AB--period. We attribute the effect to the change of the coupling $\Gamma$ as the orbital levels in the ring are tuned with a perpendicular magnetic field.
In addition, asymmetric gate voltages allow us to induce crossings of orbital levels with different symmetry. A split ZBA is observed that can be tuned using gate voltages and a in--plane magnetic field. 

\begin{acknowledgments}
We thank P. W\"olfle and K. Kobayashi for useful discussions.
Financial support by the Swiss Science Foundation (Schweizerischer
Nationalfonds) is gratefully acknowledged. 
\end{acknowledgments}

\bibliographystyle{apsrev}% Produces the bibliography via BibTeX. apsrev?

\end{document}